\documentclass[lettersize,journal]{IEEEtran}
\usepackage{algorithmic}
\usepackage[caption=false,font=normalsize,labelfont=sf,textfont=sf]{subfig}
\usepackage{textcomp}
\usepackage{graphicx}
\usepackage{cite}
\usepackage{amsmath,amssymb,amsfonts}
\usepackage{color}

\begin{document}
\title{A Digital Twin Approach for Adaptive Compliance in Cyber-Physical Systems: Case of Smart Warehouse Logistics}

% TEMPLATE
% ---------
%\author{Michael~Shell,~\IEEEmembership{Member,~IEEE,}
%        John~Doe,~\IEEEmembership{Fellow,~OSA,}
%        and~Jane~Doe,~\IEEEmembership{Life~Fellow,~IEEE}% <-this % stops a space
%\thanks{M. Shell was with the Department
%of Electrical and Computer Engineering, Georgia Institute of Technology, Atlanta,

\author{Nan~Zhang,
        Rami~Bahsoon,
        Nikos~Tziritas,
        and~Georgios~Theodoropoulos*% <-this % stops a space
\thanks{*Corresponding author.}% <-this % stops a space
\thanks{N. Zhang is with the Department of Computer Science and Engineering, Southern University of Science and Technology (SUSTech), Shenzhen, China, and also with the School of Computer Science, University of Birmingham, Birmingham, United Kingdom.}% <-this % stops a space
\thanks{R. Bahsoon is with the School of Computer Science, University of Birmingham, Birmingham, United Kingdom.}
\thanks{N. Tziritas is with the Department of Informatics and Telecommunications, University of Thessaly, Greece.}% <-this % stops a space
\thanks{G. Theodoropoulos is with the Department of Computer Science and Engineering, Southern University of Science and Technology (SUSTech), Shenzhen, China, and also with the Research Institute for Trustworthy Autonomous Systems, Shenzhen, China.}% <-this % stops a space
\thanks{This research was supported by: Shenzhen Science and Technology Program,  China (No. GJHZ20210705141807022); SUSTech-University of Birmingham Collaborative PhD Programme; Guangdong Province Innovative and Entrepreneurial Team Programme, China (No. 2017ZT07X386); SUSTech Research Institute for Trustworthy Autonomous Systems, China; and EPSRC/EverythingConnected Network project on Novel Cognitive Digital Twins for Compliance, UK.}% <-this % stops a space
%\thanks{Manuscript received April 19, 2021; revised August 16, 2021.}
}


\markboth{IEEE TRANSACTIONS ON SYSTEMS, MAN, AND CYBERNETICS: SYSTEMS}%
{Shell \MakeLowercase{\textit{et al.}}: A Sample Article Using IEEEtran.cls for IEEE Journals}

%\IEEEpubid{0000--0000/00\$00.00~\copyright~2023 IEEE}

% -------------------
% Journal Comparison:
% -------------------
% 1. {IEEE TRANSACTIONS ON SYSTEMS, MAN, AND CYBERNETICS: SYSTEMS}
% https://www.ieeesmc.org/publications/transactions-on-smc-systems/
% The scope of the IEEE Transactions on Systems, Man, and Cybernetics: Systems includes the field of systems engineering. It also includes issue formulation, analysis and modeling, decision making, and issue interpretation for any of the systems engineering lifecycle phases associated with the definition, development, and deployment of large systems. Other topics include systems management, systems engineering processes, and a variety of systems engineering methods such as optimization, modeling and simulation.
%
% 2. {IEEE TRANSACTIONS ON CYBERNETICS} 
% https://www.ieeesmc.org/publications/transactions-on-cybernetics/
% The scope of the IEEE Transactions on Cybernetics includes computational approaches to the field of cybernetics. Specifically, the transactions welcomes papers on communication and control across machines or between machine, human, and organizations. The scope includes such areas as computational intelligence, computer vision, neural networks, genetic algorithms, machine learning, fuzzy systems, cognitive systems, decision making, and robotics, to the extent that they contribute to the theme of cybernetics or demonstrate an application of cybernetics principles.
% 

\maketitle

\begin{abstract}
Engineering regulatory compliance in complex Cyber-Physical Systems (CPS), such as smart warehouse logistics, is challenging due to the open and dynamic nature of these systems, scales, and unpredictable modes of human-robot interactions that can be best learnt at runtime. 
Traditional offline approaches for engineering compliance often involve modelling at a higher, more abstract level (e.g. using languages like SysML).
These abstract models only support analysis in offline-designed and simplified scenarios. However, open and complex systems may be unpredictable, and their behaviours are difficult to be fully captured by abstract models. These systems may also involve other business goals, possibly conflicting with regulatory compliance.
To overcome these challenges, fine-grained simulation models are promising to complement abstract models and support accurate runtime predictions and performance evaluation with trade-off analysis.
The novel contribution of this work is a Digital Twin-oriented architecture for adaptive compliance leveraging abstract goal modelling, fine-grained agent-based modelling and runtime simulation for managing compliance trade-offs. 
A case study from smart warehouse logistics is used to demonstrate the approach considering safety and productivity trade-offs.   

\end{abstract}


\section{Introduction}
\label{sec:introduction}
%The narrative: legal interpretation leads to different design alternatives; these alternatives can all satisfy the legal compliance; but they lead to different satisfaction of system goals; For a highly dynamic environment, it is difficult to enumerate all possible situations during design time; there is a need to allow adaptive switching between different 
% Background on compliance 
\IEEEPARstart{R}{egulatory} compliance is ``the act of ensuring adherence of an organisation, process or (software) product to laws, guidelines, specifications and regulations'' \cite{akhigbe_systematic_2019}.
This paper specifically focuses on the compliance for software products.
Emerging digitalisation trends in application domains like smart manufacturing and smart logistics emphasise the necessity of compliance with various regulations and standards (e.g., General Data Protection Regulation [GDPR], ISO series, safety, security, etc.) \cite{kumar_survey_2021,yue_smart_2018}.
These systems may involve a large number of smart entities, such as swarms of autonomous mobile robots, internet-of-things, etc., which poses new requirements for the design of the system in compliance with standards such as safety \cite{iso_3691-42020_industrial_2020}.  
These systems are essentially complex Cyber-Physical Systems (CPS) in which information from the physical world is continuously collected to the cyber world, and computation, communication and control enabled by software (the cyber part) manage the behaviour of the physical systems (the physical part) in return to perform dedicated tasks autonomously and collaboratively \cite{radhakisan_baheti_cyber-physical_2011}.
These highly complex CPS are systems of systems, and often operate in open and dynamic environments, facing unpredictable behaviours and sometimes involving interaction with humans, making it a challenge to realise the desired level of compliance (e.g., safety). 

% We focus on compliance for the product in service time/runtime.
% For autonomous CPS, we need to design software that enables compliant autonomous behaviours at runtime.
\subsection{Challenges} \label{sec: challenges}
Autonomy is an important dimension in CPS that enables the system to perform specific tasks in dynamic and complex environments without human intervention \cite{muccini_self-adaptation_2016,bures_software_2017,leitao_smart_2016}. Compliance with regulations such as safety needs to be guaranteed for autonomous behaviours in the face of different environments at runtime. These autonomous behaviours are enabled by software, which poses requirements for designing software that enables safety compliance behaviours at runtime. 
Engineers need to design the software-enabled communication, computation and control scheme with the desired level of compliance in mind to ensure acceptance. For instance, in human-robot collaboration, safety is a major concern. 
Automatic motion control strategies need to ensure that humans are safe while working alongside robots \cite{liu_human-robot_2019}.

%\IEEEpubidadjcol

% Challenges for ensuring compliance for autonomous CPS
Designing for compliance requires the mapping from compliance sources (specifications of regulations, standards, etc.) to software-regulated behaviours. 
However, for autonomous CPS in a complex dynamic environment, the following challenges may arise.% in the analysis and governance of compliance.

% Problems at runtime: trade-off between compliance and other goals
\paragraph{Compliance and quality goals}
CPS is designed to perform specific tasks in the physical environment. In addition to complying with regulations, the system aims to achieve other system-level or business goals.
One essential aspect is maximising the productivity of the CPS, such as improving package delivery efficiency in an automated warehouse. 
Nonetheless, conflicts often arise between meeting compliance requirements and pursuing other goals such as productivity, profit, and cost \cite{van_asselt_simulating_2016}. A trade-off analysis is necessary.

% limited knowledge, time, budget
\paragraph{Limited design effort and knowledge}
%Mapping compliance sources to the product or process is constrained by limited or biased contextual knowledge, time, budget, priorities, etc.
Quality assurance engineers often take a ``best effort'' approach to map standards, guides and regulations to process, control, and infrastructure. However, this effort of mapping is often constrained by time, budget, and priorities, and is often biased to the expertise of the analysts. Additionally, the mapping process is complicated by considering various dependencies and conflicts between standards, guides and regulations, making it difficult to ensure optimal compliance at design time. Therefore, this difficulty calls for dynamic and continuous evaluation approaches for compliance, considering interaction trends (whether direct or implied) of various CPS components and their interactions.

% Problems at runtime: cannot foresee all possibilities at design time
\paragraph{Dynamic runtime behaviour}
% Dynamic complex runtime environment that is not easily predicted during design time, because of problems such as state explosion, etc...
For complex CPS, testing compliance for the software-controlled behaviour at design time can suffer from incompleteness and inadequacies. 
The designed control policies may only be applicable in specific scenarios, since not all possible future threats can be foreseen during the design phase and the full behaviour of the complex environment only emerges during operation \cite{sanchez_survey_2019,vierhauser_requirements_2016}. 
CPS are usually systems of systems, and the interaction between multiple sub-systems increases the complexity of the overall system behaviour. Additionally, the existence of human and machine-related errors and their ripple impact on the system can best be understood at run-time due to the absence of necessary contextual information during design time. 
% Therefore we need runtime governance
Henceforth, runtime compliance monitoring becomes a necessity for the sustainable and dependable operation of the CPS, and as a mitigation strategy for preventing degradation in the system and avoiding Service Level Agreement (SLA) violations.  

% Problems at runtime: legal interpretation -> different design alternatives -> different satisfaction of goals
\paragraph{Legal interpretation}
Legal and regulatory documents tend to be ambiguous and describe general norms, which sometimes need interpretation \cite{massey_identifying_2014,ghanavati_impact_2015}. 
As mentioned by Sánchez et al, ``it is hard for engineers to assess and provide evidence of whether a technical design is compliant with the law due to the gap existing between a legal document written in natural language and a technical solution in the form of a software system'' \cite{sanchez_survey_2019}.
In achieving a certain regulatory/standard compliance, different interpretations can lead to different design alternatives of the software product, which impact differently on other quality goals, also calling for trade-off analysis \cite{ghanavati_impact_2015,sartoli_compliance_2020}.
Due to the dynamism of the environment,
control policies that were optimal in the past may later become obsolete, thus decreasing the satisfaction of quality goals in new unforeseen environmental contexts. Strategies that are optimal for a certain part of the system may also lead to compliance violations and goal degradation in other parts of the system.

The composite effect of the above challenges makes it difficult for traditional offline design methodologies to handle the uncertainties only known at runtime. It is difficult for system testing at design time to foresee or enumerate all possible environmental contexts.
Therefore, the analysis and governance of compliance for the autonomous software-regulated behaviour of CPS should be extended from design time to runtime in an adaptive manner with dynamic trade-off analysis.
%Trade-off analysis is needed at runtime for different control strategy alternatives, selecting the strategy that maximises productivity goals while remaining compliant in different environmental contexts. 

%(To summarise) In achieving a certain regulatory/standard compliance, there can be different design alternatives of the compliance control, which impact differently on other system goals, calling for trade-off analysis \cite{sartoli_compliance_2020}

% Therefore, Continuous monitoring of the compliance status and adapting strategy at runtime for the remedy of system degradation are needed.

% Some existing solutions
Research in compliance has mostly been considered offline, with the investigation for runtime adaptive compliance emerging \cite{garcia-galan_towards_2016}. 
However, the modelling and analysis of adaptive compliance are mostly at higher abstraction levels, such as goals and with tools such as SysML \cite{akhigbe_systematic_2019,sartoli_compliance_2020,anda_modeling_2018}.
Models at this level of abstraction alone may fail to support analysis for complex and open environments, whose behaviour can involve cascading effects due to the interaction between sub-systems. The behaviour of these environments can only be best anticipated by simulation models that capture the fine-grained detailed information about each sub-system. 
Also, there is a lack of extensive studies for adaptive compliance in a real-time fast-changing environment. 

\subsection{Contribution}

Digital Twin (DT) can be a promising solution to the above challenges in real-time compliance governance for autonomous CPS in highly dynamic environments. The concept of DT stems from manufacturing and is now being applied to various application domains \cite{tao_digital_2019_8049520,glaessgen_digital_2012,kuhl_warehouse_2022,ayyaz_aquaponics_2019}.
DT contains high-precision model replicas of the real system/environment \cite{thelen_comprehensive_2022}. The core of DT is Dynamic Data Driven Applications Systems (DDDAS), enabling real-time data sensed from the system to be continuously assimilated into the model to update the model for better fidelity and to inform more accurate runtime prescriptive what-if analysis and decision-making, which in turn affects the behaviour of the running system \cite{Zhang2020}.
%\TODO{Although there have been attempts in combining compliance model and system models like SysML for adaptive socio-cyber-physical systems \cite{anda_modeling_2018}}, 
Research is still in its infancy in utilising DT's high-fidelity simulation model to support analysis based on fine-grained behaviour predictions and runtime what-if analysis.

This paper specifically aims at applying DT for adaptive runtime compliance governance for the autonomous software-controlled behaviours of systems of systems with human-in-the-loop.
To the best of our knowledge, no existing work uses faster-than-real-time \textit{computation} supported by a continuously updated fine-grained simulation model (i.e. DT) for the runtime governance of compliance.
There is still a lack of extensive investigation for the runtime feedback loop in the compliance context that enables autonomous model update, reasoning and self-adaptation.
In particular, we make the following novel contributions: 
\begin{itemize}
    \item A Digital Twin-based architecture for runtime compliance governance that incorporates abstract goal modelling and detailed agent-based modelling for runtime monitoring, what-if analysis and adaptive control.
    \item The utilisation of goal modelling for compliance especially in how its refinement is linked to the design alternatives of agents.
    \item The runtime trade-off analysis for design alternatives on their satisfaction of compliance and other quality goals with Pareto efficiency. 
    \item The applicability of the approach is demonstrated in a case study of human-robot collaboration in smart warehouse logistics with dynamic trade-offs for safety compliance, production, and workforce constraints. The results show that Digital Twin can provide engineers with insights for live evaluation and continuous refinements of the system's behavioural rules for better compliance.  
\end{itemize}

%\NEW{We scope the focus on designing software of autonomous CPS that enables compliant runtime behaviours.}
%\NEW{We focus on the adaptive runtime governance of the trade-off of compliance and other quality goals.}
%\NEW{We focus on the design alternatives generated by legal interpretation.}
%\NEW{We combine abstract modelling (goal modelling) and detailed simulation models (agent-based modelling)}

%\NEW{human-in-the-loop}
%added values

The rest of the paper is organised as follows. Section \ref{sec:related_work} provides the related work in adaptive compliance. Section \ref{sec:problem} formulates the problem of adaptive compliance in the smart warehouse context. Then the DT solution is proposed in Section \ref{sec:DT_solution}. Section \ref{sec:case_study_and_modelling} describes the case study of the smart warehouse and the detailed compliance modelling adopted by this paper. The experimental evaluation is presented in Section \ref{sec:evaluation}. Finally, this paper concludes in Section \ref{sec:conclusion}.

%\TODO{
%Mostly focus on higher level such as business or software process, not simulation at the system level.
%}
%\TODO{
%The gap:
%The usage of Digital Twin especially info-symbiotic dynamic %data-driven simulation at the system level is lacking.
%}

%\TODO{
%How to evaluate legally interpreted design at runtime and %dynamically apply them.
%}

%\TODO{
%Our contribution:
%Incorporate goal modelling and system level simulation especially agent-based models.
%Monitoring goal satisfaction.
%Runtime dynamically evaluating and changing design alternatives out of compliance interpretation using system-level digital twin simulation.
%}

\section{Related Work} \label{sec:related_work}

This section presents the related work in runtime compliance governance in business processes and requirements engineering, as well as the initiative of adaptive compliance. 

\subsection{Runtime Compliance Governance for Business Process}
Compliance has been extensively discussed in business domains, especially in process-oriented business environments. Comprehensive systematic literature reviews can be found in \cite{hashmi_are_2018} and \cite{mustapha_systematic_2020}.
%There are five phases in the life-cycle of compliance management: elicitation, formalisation, implementation, checking/analysis, and optimisation \cite{ramezani_separating_2012}. 
%Suitable technologies and approaches are needed to support each phase, especially in the emerging context of digitalisation such as cyber-physical systems \cite{sackmann_using_2018}.
In business processes, runtime monitoring and detection of possible compliance violations are essential, since it would be infeasible at design time to anticipate the satisfaction of runtime aspects such as timing and resource assignment constraints \cite{barnawi_runtime_2015}.
Through interpretation, compliance requirements can be transformed into objectives and subsequent specifications as compliance rules or constraints. Then the monitoring can be made possible against the rules and constraints \cite{ly_compliance_2015}.

Many existing efforts have been made in designing architectural frameworks for runtime compliance monitoring and governance. For instance, \cite{birukou_integrated_2010} presented an integrated framework using Service-Oriented Architectures (SOAs) to support the whole compliance management lifecycle, including modelling, monitoring, displaying and reporting on violations.
Various Domain-Specific Languages (DSL) and Business Process Execution Language (BPEL) are used in the modelling.
%Various Domain-Specific Languages (DSL) are used to model compliance requirements and Business Process Execution Language (BPEL) is applied to model business processes. 
%\TODO{However, no process re-design is involved in the framework}
In \cite{barnawi_runtime_2015}, the authors propose a runtime self-monitoring approach to business process compliance in cloud environments. They embed the monitoring logic within the process model without having an external monitoring component.
%designed for augmenting the Business Process Modeling Notation (BPMN) model.

Runtime compliance monitoring is also enabled by specific languages for modelling compliance rules such as Compliance Rule Graphs (CRG) and extended CRG (eCRG) \cite{ly_monitoring_2011,knuplesch_framework_2017}.
In \cite{knuplesch_framework_2017}, visual compliance monitoring is made possible with eCRG to support violation detection and cause traceability.

Beyond the specific monitoring approaches, \cite{ly_compliance_2015} proposed a systematic conceptual framework for comparing approaches for compliance monitoring. Ten compliance monitoring functionalities (CMF) are elicited that can characterise the capability of any compliance monitoring approach.

Compliance in business processes is related to regulating human behaviours in an organisation to follow the rules. In contrast, this paper focuses on designing 
software-controlled autonomous behaviours that follow regulatory requirements.

\subsection{Compliance and Requirements Engineering}
Compliance for software products is commonly discussed and analysed in requirements engineering \cite{raykar_iterative_2021,breaux_towards_2006}. 
Requirements engineers need to extract relevant requirements (e.g. rights and obligations) from the legal text and monitor the compliance of the software in its whole lifecycle \cite{breaux_towards_2006,otto_addressing_2007}.
Such monitoring is also essential for complex software systems of systems \cite{vierhauser_requirements_2016}.  

% Modelling and analysis of compliance
The modelling of regulations from a requirements engineering perspective includes using logic, goal models, and semi-structured representations \cite{otto_addressing_2007}.
Among the approaches, goal modelling is generic for modelling and analysing software requirements and can also be applied to compliance modelling. Goal models can represent not only the intent but also the structure of law \cite{akhigbe_systematic_2019}.
For instance, the work \cite{ingolfo_arguing_2013} utilises \textit{i*} and \textit{Nòmos} to express the requirements and regulations, where stakeholders' inputs can be used to refine the compliance model. 
The work in \cite{negri-ribalta_socio-technical_2022} proposes an extended version of a goal- and actor-oriented modelling language called \textit{STS-ml} tailored for GDPR-specific principles and concerns.
The work in \cite{ojameruaye_systematic_2014} proposed the concept of compliance debt as a kind of technical debt.
Neglecting or not imposing compliance is modelled as a kind of compliance debt, and the debt is used as a decision factor. Combined with economics-driven approaches, they manage the risk of tolerating these obstacles that impede the satisfaction of compliance requirements.
%Tolerance of obstacles is a kind of compliance debt, and needs to be analysed against its risk.
In addition, extracting requirements from legal documents can be automated by deep learning large language models \cite{sainani_extracting_2020,abualhaija_automated_2022}.

This paper adopts goal modelling, a modelling approach from requirements engineering to support requirements elicitation and runtime analysis of design alternatives in the DT.

\subsection{Adaptive Compliance}
Modern CPS involves highly configurable software and operates in dynamic environments, which poses new requirements for self-adaptive compliance management at runtime.  
Adaptive compliance in software-based systems has been motivated by García-Galán et al., who apply the Monitor-Analyse-Plan-Execute (MAPE) loop for runtime self-adaptation towards continuous satisfaction of compliance requirements in case of any variabilities \cite{garcia-galan_towards_2016}. 
Violation of compliance can come from the variability of compliance sources, the systems, and the operational environment \cite{garcia-galan_towards_2016}.
There are some preliminary studies in adaptive compliance. For instance, the work in \cite{sartoli_compliance_2020} uses Goal-oriented Requirements Language (GRL) to model the variability of the environment in relation to sources. Different design alternatives for compliance control can exist, each with its trade-offs \cite{sartoli_compliance_2020}.
However, there is still a lack of using detailed computational modelling for behaviour simulation prediction to tackle the challenge of runtime compliance management for complex systems of systems.

%``designing a compliance-monitoring framework for detection of data-handling violations in real time.'' \cite{happa_run-time_2019}

One perspective in supporting self-adaptive behaviour is \textit{Models@Run.Time}. Runtime models are self-representations of behaviour, goal, and structure of the system: changes made on the model at runtime can be reflected in the real system, and vice versa \cite{bencomo_modelsruntime_2019}.
Formal methods such as real-time model checking \cite{nishizaki_real-time_2013} and runtime verification \cite{sanchez_survey_2019} are common approaches for monitoring and verifying possible specification requirement violations based on the runtime model.
However, formal methods suffer from scalability issues such as state explosion problems. Models in formal methods usually apply to closed systems whose total number of states is relatively size-limited. Applying these methods to open systems such as multi-agent cyber-physical environments can be challenging.

\section{Problem Formulation} \label{sec:problem}
To address the aforementioned challenges of section \ref{sec: challenges}, this paper adopts a scenario of smart warehouse logistics and formulates the problems that arise.
The questions to be investigated are: how a DT can overcome these challenges for the smart warehouse scenario? And how to model the problem and system using the DT approach? 

\begin{figure}[!t]
\centerline{\includegraphics[width=\columnwidth]{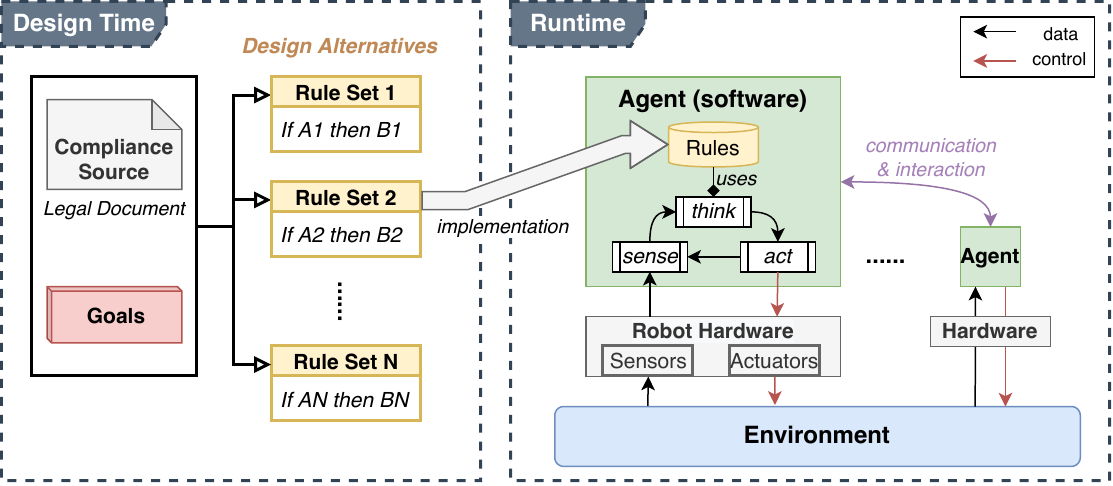}}
\caption{Design process: from compliance source to behavioural rules of agents.}
\label{fig_physical_space}
\end{figure}

% A brief setup of the warehouse scenario: intelligent robots, human-robot collaboration
The smart warehouse in this paper is considered a multi-agent system consisting of Autonomous Mobile Robots (AMRs), collaborative robots (cobots), and human workers.
Such a system is an autonomous complex system, with components (e.g., robots, humans, IoT sensors and intelligent processors, etc.) collaboratively working with each other or/and with humans to conduct logistics-related tasks. Each robot in the system is controlled by an onboard intelligent software agent that follows a \textit{sense-think-act} cycle, as shown in the right part of Fig. \ref{fig_physical_space}.
The agent uses sensors mounted on the robot to continuously perceive its surrounding environment, including information about other robots and humans, and tasks in the pipeline.
In this paper, each agent is assumed to follow specific behavioural rules designed to adhere to compliance requirements.
One common type of compliance is safety, which requires agent-controlled AMRs to slow down or stop when human workers are nearby.

The challenge is to devise an adaptive tuning mechanism of the agents' behavioural rules under a highly dynamic environment, such that the smart warehouse system can 1) ensure regulatory compliance, such as safety while 2) maximising the satisfaction of quality goals such as productivity.
The two objectives are usually conflicting, since a robot that is too safe may stop frequently even when a human worker is detected in its surroundings but far away from it, thus delaying the delivery of the robot's loaded package to the destination.
In addition, as shown in Fig. \ref{fig_physical_space}, there can be multiple design alternatives for the rules of agents in the system, some of which can result from subjective interpretation of compliance sources.
As discussed earlier, it is difficult for these design alternatives to be evaluated offline with full credibility, since compliance engineers are often challenged by limited knowledge, and possibly inflated, deflated or wrong assumptions at design time, especially for systems with novel situational and contextual use and new application domains. Additionally, some rules might not necessarily render compliance for all run-time scenarios, and contextualisation of these rules might be necessary to reach the desirable level of compliance.
Consequently, it is imperative that behaviour rules for compliance need to be dynamically adapted at runtime to better cater for different scenarios and optimise for the trade-off between compliance and quality goals. 

This paper proposes to use DT to tackle the challenge of the runtime trade-off between compliance governance and quality goals for complex systems of systems that operate in dynamic environments. 
In particular, we aim to investigate the following questions: 
1) how to utilise the DT approach to model the problem of runtime adaptive trade-off analysis?
2) How DT can better explore the design space, and better understand the rules in different contexts?

% How DT can better understand the rules
% What DT can offer, that is difficult using traditional approahces
% How DT can better explore the design space, and better understand the trade-off, dynamic trade-off
% knowledge got at runtime, can be different from that at design time.
% link to the challenges 

\section{Digital Twin for Adaptive Compliance} \label{sec:DT_solution}
Digital Twin (DT) is a paradigm that utilises high-fidelity computational modelling to create virtual replicas of real-world entities in the virtual world. DT can leverage the principle of Dynamic Data-Driven Applications System (DDDAS) to utilise two-way communication between the real world and the virtual world to update its simulation model continuously and to provide simulation-informed decisions and control back to the real world \cite{Zhang2020,darema_dynamic_2004}.
This section presents our DT solution for runtime compliance governance considering the trade-off with productivity goals, including the reference architecture, modelling and control.

\subsection{A Reference Architectural Framework}
\begin{figure}[!t]
\centerline{\includegraphics[width=\columnwidth]{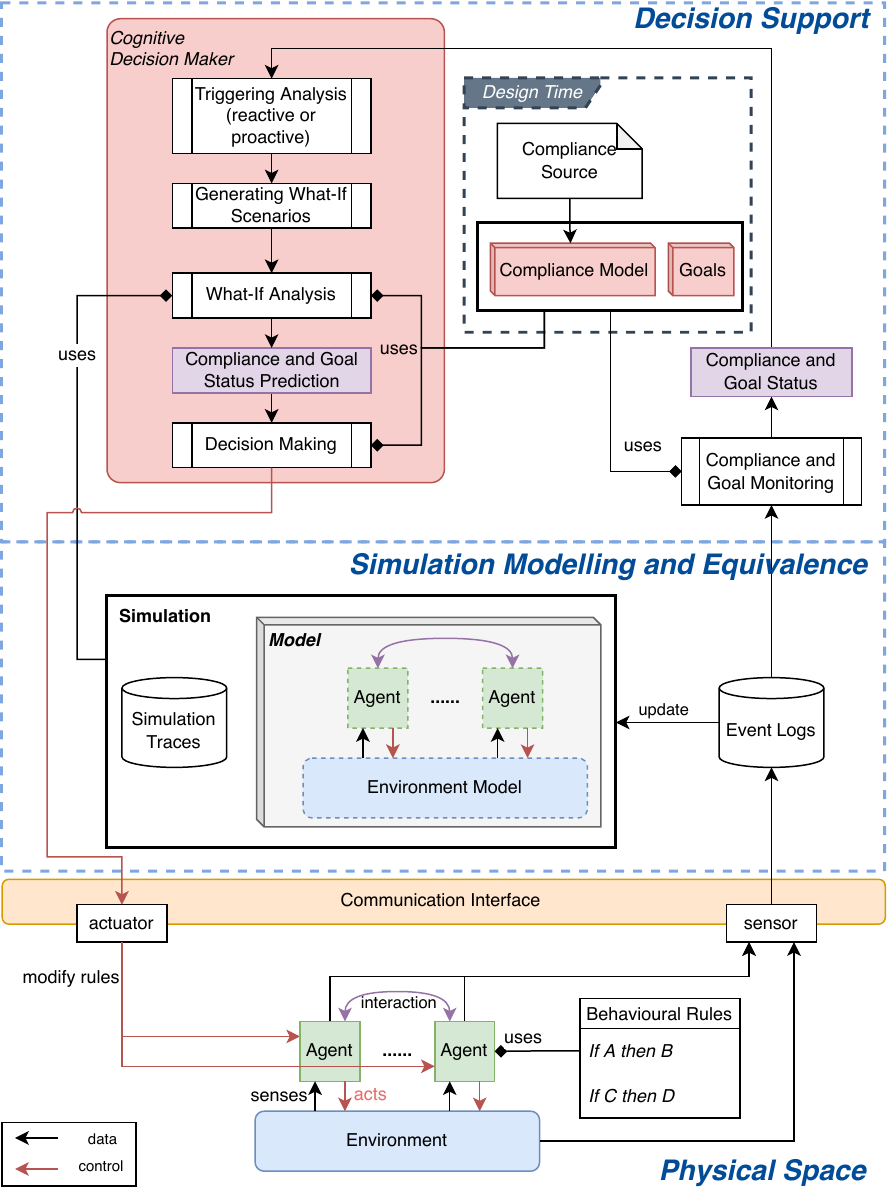}}
\caption{The reference architecture for a compliance-aware Digital Twin.}
\label{fig_arch}
\end{figure}

This subsection describes a novel DT-oriented reference architecture for adaptive compliance, which builds on the foundation set in our previous work to serve the compliance case \cite{zhang_knowledge_2022}. 
The architecture is shown in Fig. \ref{fig_arch}, which mainly contains three parts: \textit{physical space}, \textit{simulation modelling and equivalence}, and \textit{decision support}. The latter two parts constitute a novel extension to realise intelligent compliance-aware DT.

The \textit{physical space} contains robots (and their onboard controlling agents), human workers, and all other industrial assets. As mentioned before, the industrial system is viewed as a multi-agent system. Each agent has its own rules and can receive information and meta-level instructions (i.e. to adapt the behavioural rules) from the DT. 
The state and events of the environment and agents are continuously monitored by the DT in real time via various IoT sensors.

The \textit{simulation modelling and equivalence} part is depicted above the physical space. Its core is an equivalent agent-based simulation model of the physical space that can forecast the behaviour of the physical space through what-if analysis for various scenarios. The simulation model replicates the environment (including the properties of humans, robot hardware, and other industrial assets) and the agents.
The modelling of the environment can use a state-based approach where a set of state variables (e.g. locations of AMRs and humans) characterise the status of the environment at any particular moment. The evolution of the entire system model is controlled by other internal parameters (e.g. parameters in the behaviour rules of AMRs).
For the modelling of agents, since agents are essentially software programs, the ``twinning'' of the agents in the real-world AMRs can be achieved by replicating the same software program into the simulation model.
The physical system's state and events are monitored and continuously assimilated into the model to perfect its fidelity for prediction.
The agent-based model in the DT is updated in real-time in a data-driven manner through state replication and model calibration of the internal parameters. 

The \textit{decision support} part provides runtime analysis and optimisation for compliance and quality goals.
This part monitors the status of compliance and goal satisfaction and informs further analysis.
A cognitive decision maker is introduced; this novel addition utilises the up-to-date simulation model to run multiple what-if analysis at runtime by simulating the outcome of different design alternatives (as shown in the left part of Fig. \ref{fig_physical_space}) out of the same compliance source and productivity goals given the current state of the physical space. This analysis is essential since one certain design choice of the rule set may not be optimal in all possible situations during runtime. Such runtime what-if simulation analysis enables evaluating the applicability of different modifications of the agents' rules in real-time for trade-off concerns.
The outcome evaluation also requires models that are able to represent the semantics of compliance and goals, and that can quantitatively measure the satisfaction of compliance and goals. In this paper, Goal-oriented Requirement Language (GRL) is adopted for the modelling and refining of both compliance goals and specific requirements for the considered cases. 
%The detailed discussion is presented in the next section.

Finally, the candidate rule modification (one of the design alternatives elicited at design time) that leads to better trade-offs will be enacted in the agents in the physical space. Such an adaptation is done via \textit{software} actuators of the DT only to modify the agent program.

\subsection{Compliance Modelling and Assessment}

Various approaches have been proposed for modelling compliance, among which goal modelling has been broadly studied.
The advantage of goal modelling for compliance is that it addresses both the intent and structure of the law or regulations \cite{akhigbe_systematic_2019}. 
Many of the studies relate compliance with requirement engineering such that legal requirements and other system requirements can be aligned and reasoned altogether.

Goal-oriented Requirement Language (GRL) is a standardised language for modelling the objectives of stakeholders and systems, as well as their relationships. 
It can be used to elicit compliance requirements as goals and their solutions and obstacles, and represent them graphically as a tree-like structure.
GRL supports quantitative and qualitative trade-off analysis for the satisfaction levels of the intentional elements (e.g., goals and tasks) and actors \cite{amyot_evaluating_2010}.
GRL has also been further extended with systematic guidelines for extracting and modelling legal statements \cite{ghanavati_legal_2014}.

This paper uses GRL to model the compliance requirements and other system goals at design time. The satisfaction of requirements will be quantified by defining a set of metrics. 
Through goal refinement, GRL can provide a range of design alternatives for the rules of agents.
Then the goal model and metrics will be integrated into the DT-oriented architecture and utilised at runtime. Design alternatives not implemented in the agents will be evaluated at runtime using faster-than-real-time simulation to assess their satisfaction levels based on the metrics quantitatively.

\begin{figure}[!t]
\centerline{\includegraphics[width=\columnwidth]{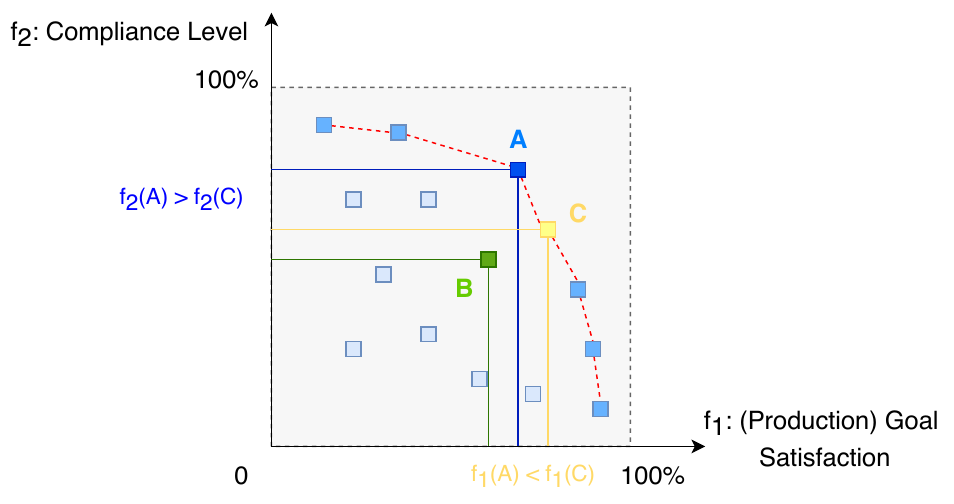}}
\caption{Pareto efficiency of two objectives: compliance level and goal satisfaction. A and C both dominate B, but A and C are mutually non-dominated solutions.}
\label{fig_pareto}
\end{figure}

Runtime assessment of these alternatives involves comparing them against multiple objectives (compliance and production goals). Pareto efficiency (or Pareto optimality) can be used for the comparison.
Pareto efficiency is based on the \textit{non-dominance} relation between solution alternatives, as shown in Fig. \ref{fig_pareto}.
A solution $x$ dominates $y$ if and only if $x$ is at least as good as $y$ in all objectives and better in at least one \cite{while_faster_2006}. A solution $x$ is \textit{non-dominated} if and only if no other solutions dominate $x$. The set of all mutually non-dominated solutions is called the \textit{Pareto front} \cite{while_faster_2006}.

%\subsection{Notations}

%\TODO{Brief introduction for GRL notations}

%Goal refinement: different design alternatives of agent’s behavioural rules.

%Legal interpretation and productivity goals

%\subsection{Rule alternatives of agents}
%\TODO{Linking GRL ``tasks'' with agent rules}

%\TODO{A figure showing alternative rules in general terms}

%\subsection{Monitoring Metrics}

\section{Case Study: Human-Robot Collaboration} \label{sec:case_study_and_modelling}
\begin{figure}[!t]
\centerline{\includegraphics[width=\columnwidth]{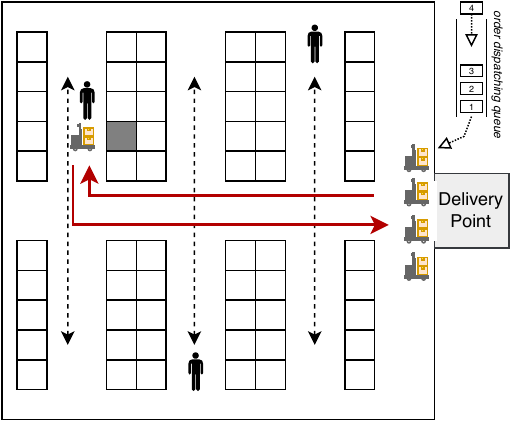}}
\caption{Overall layout of the case study. Human workers move according to black dashed arrows. The movement of AMRs (red arrows) may intersect with the workers. The safety compliance behavioural rule onboard an AMR should instruct the AMR to slow down to avoid collision with human workers.}
\label{fig_use_case}
\end{figure}

This section presents a case study of human-robot collaboration in smart warehouse logistics to demonstrate the DT approach. 
The warehouse layout is shown in Fig. \ref{fig_use_case}.
Multiple AMRs and human workers work together in the warehouse to collaboratively pick and deliver requested items from incoming orders.
Each order requests one item that is stored in one slot on a rack in the warehouse. Orders first arrive in a central dispatching system, and wait in queue until there is any idle AMR. When at least one AMR is available, the dispatching system will assign one order to one AMR.
The AMRs are assumed not to be specialised in picking items off the racks, but only designed for ground transportation. For this reason, human workers will help in picking the items off the racks and loading the items onto the AMRs. 

The behaviour workflow of the AMRs and human workers is as follows. 
The AMRs are initially in an idle state and waiting for orders. After receiving the order information, each AMR needs to first plan its path to the location of the item. Then the AMR notifies a human worker near that location by sending messages to the worker's smartphone. The AMR and the worker move to the location and wait for each other to arrive. The worker picks the item off the rack and loads it onto the AMR. Finally, the AMR carries the item and returns to the delivery point to finish delivery and then becomes idle.

For such a semi-automated warehouse, the number of human workers should be minimised to reduce the labour cost. This means that a limited number of human workers need to move to different locations in the warehouse to assist the AMRs, thus posing safety compliance requirements on the design of the AMR's autonomous movement control.

\subsection{Compliance Modelling with GRL}
We focus on safety compliance as a concern for this case study. We also consider how this concern poses a trade-off against the productivity goal considering spatial requirements for safety and human-robot collaboration (individual and collective). 

The safety compliance concern can be illustrated by Fig. \ref{fig_use_case}. Workers need to move to different places within the warehouse, which may easily intersect with the paths of AMRs, causing collision and possible injury. 
Compliance with safety requirements for driverless industrial trucks has been specified in ISO 3691-4:2020, in which the \textit{collision with person} is one of the main risks \cite{iso_3691-42020_industrial_2020}. 
Specifically, in ISO 3691-4:2020 clause 4.8.2.1, detection of persons and collision avoidance are required:
\textit{``c) Personnel detection means shall be so designed that trucks shall stop before contact between the rigid parts of the truck or load and a stationary person (not a person stepping into the truck path or moving toward it) [...]''}
In addition, such detection can be adaptive according to clause 4.8.2.6 (Selection of the active detection zone fields):
\textit{``Trucks can have an automatic selection of the safe detection fields based on truck speed and direction, size of the load or other criteria [...]''}
It can be seen that the above requirements specified in the ISO document are ambiguous and general, which can be interpreted into different design alternatives and choices for the control in AMRs.

\begin{figure}[!t]
\centerline{\includegraphics[width=\columnwidth]{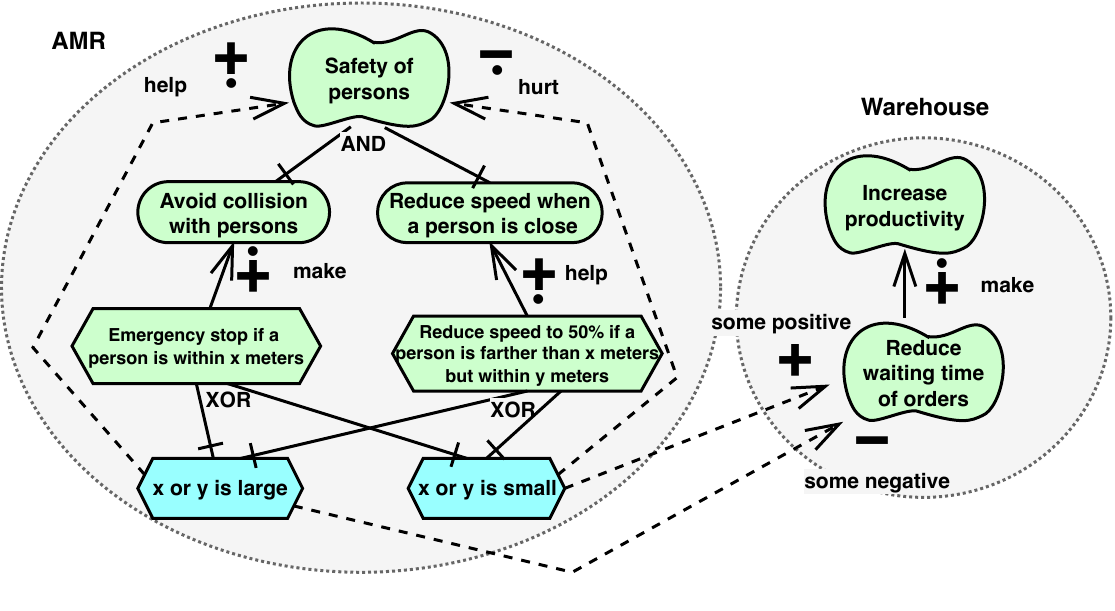}}
\caption{The GRL model for safety compliance and productivity.}
\label{fig_goal}
\end{figure}

The GRL model of both compliance and production goals is shown in Fig. \ref{fig_goal}. 
The production goal is denoted as ``\textsf{increase productivity}'' and further refined as ``\textsf{reduce waiting time of orders}''.
Safety compliance is regarded as a softgoal or non-functional requirement for each AMR, denoted as ``\textsf{safety of persons}'', which is further decomposed into two sub-goals: avoid collision and reduce speed.
These two sub-goals specify two safety strategies to ensure the safety of the persons. They are connected to the top-level goal with an \textsf{AND} operator, meaning both two safety mechanisms need to be implemented. 
Each sub-goal is provided with a solution.  
The goal of collision avoidance can be sufficiently achieved by implementing a behavioural rule for emergency stops in the AMR, which is required by ISO 3691-4:2020 clause 4.8.2.1.
For the sub-goal of speed reduction, it adds an extra level of safety without stopping the AMR's delivery task. 
In achieving the sub-goal of speed reduction, one solution is to design a rule for speed control with distance criteria, which is shown as a task element in Fig. \ref{fig_goal}: ``\textsf{reduce speed to 50\% if a person is farther than x meters but within y meters}''. This task contributes to the satisfaction of its higher-level goal of speed reduction.

The distance criteria x and y in the two tasks (the two green hexagons in Fig. \ref{fig_goal}) can inform various design alternatives. 
Two examples of design alternatives are shown as sub-tasks in Fig. \ref{fig_goal} denoted by cyan hexagons, indicating whether to use small or large values for the criteria x and y. The label \textsf{XOR} means only one of the two design alternatives will be chosen.
These alternatives affect differently on the satisfaction of the compliance goal and the production goal, leading to a trade-off analysis.
For instance, with a larger criterion, an AMR is safety compliant but may reduce its speed even when a person is too far away. In a busy scenario where a large number of human workers are involved, the AMR may reduce its speed too frequently such that its delivery can be delayed.

\subsection{Metrics for Goal Satisfaction}

We then design metrics to measure the levels of satisfaction of the two goals in the GRL model.

For safety compliance, first, a safety value is calculated for each person. Then the overall level of safety compliance can be measured by the mean safety value over all persons.
The safety value of a person $i$ is proportional to the distance between the person and the nearest AMR moving towards him/her. This distance is denoted as $d_i(t)$.
Then the safety value of the person $i$ is:
\begin{equation}
    s_i(t) = \frac{d_i(t)}{\max{(d_i(t), d_{th})}}
\end{equation}

where $d_{th}$ is a threshold: if $d_i(t) > d_{th}$, then the person is regarded as 100\% safe meaning the AMRs are all too far away.

Suppose the warehouse involves in total $N$ human workers.
The overall safety level can be given in two ways using the mean or minimum value:
\begin{equation}
    \text{Safety}_{\text{mean}}(t) = \frac{\sum_{i=1}^{N}{s_i(t)}}{N}
\end{equation}

\begin{equation} \label{eqn: safety_min}
    \text{Safety}_{\text{min}}(t) = \min_i{\{s_i(t)\}}
\end{equation}
%\begin{equation}
%    s_i(t) = \frac{\sum_{j \in R}{d_{ij}(t) \cdot w_{ij}(t)}}{\sum_{j \in R}{w_{ij}(t)}}
%\end{equation}
%\begin{equation}
%    w_{ij}(t) = \frac{1}{d_{ij}(t)}
%\end{equation}
%\begin{equation}
%    s_i(t) = \frac{card(R)}{\sum_{j \in R}{\frac{1}{d_{ij}(t)}}}
%\end{equation}

For productivity, the average service time of past $n$ completed orders is used to measure the performance.
Suppose at time $t$, the most recently completed order is the $k$-th one counted from the initial starting of the smart warehouse system, then the productivity is measured as:
\begin{equation}
    \text{AvgServiceTime}(t,n) = \frac{\sum_{i=k-n+1}^{k}{(T^i_{\text{completion}} - T^i_{\text{arrival}})}}{n}
\end{equation}

\begin{equation}
    \text{Productivity}(t) = 1 - \frac{\text{AvgServiceTime}(t,n)}{\max{(T_{th},\text{AvgServiceTime}(t,n))}}
\end{equation}

where 
\begin{itemize}
    \item $T^i_{\text{arrival}}$ is the time when the $i$-th order arrives and starts to wait in the dispatching queue.
    \item $T^i_{\text{completion}}$ is the completion time of the $i$-th order, i.e. when the requested item is transported to the delivery point.
    \item $T_{th}$ is a threshold value that specifies the largest tolerable service time.
\end{itemize}

The overall satisfaction of the two goals can be evaluated according to Pareto efficiency or a weighted sum:
\begin{equation}
    \text{Overall}(t) = w_s \cdot \text{Safety}(t) + w_p \cdot \text{Productivity}(t)
\end{equation}

where $w_s + w_p = 1; w_s, w_p \in [0,1]$.

\section{Experimental Evaluation} \label{sec:evaluation}

We aim to evaluate our solution by demonstrating the necessity and advantages of DT \textit{runtime} analysis compared to offline analysis.
As stated in Section \ref{sec:problem}: how can DT better explore the design space, and better understand the rules in dynamic changing contexts or scenarios?
The experiment will evaluate DT-enabled analysis in changing environment and its ability to obtain insights in those scenarios.

% 1) how to utilise the DT approach to model the problem of runtime adaptive trade-off analysis? 
% 2) How DT can better explore the design space, and better understand the rules in different contexts?

\subsection{Experiment Setup}

\begin{figure}[!t]
\centerline{\includegraphics[width=\columnwidth]{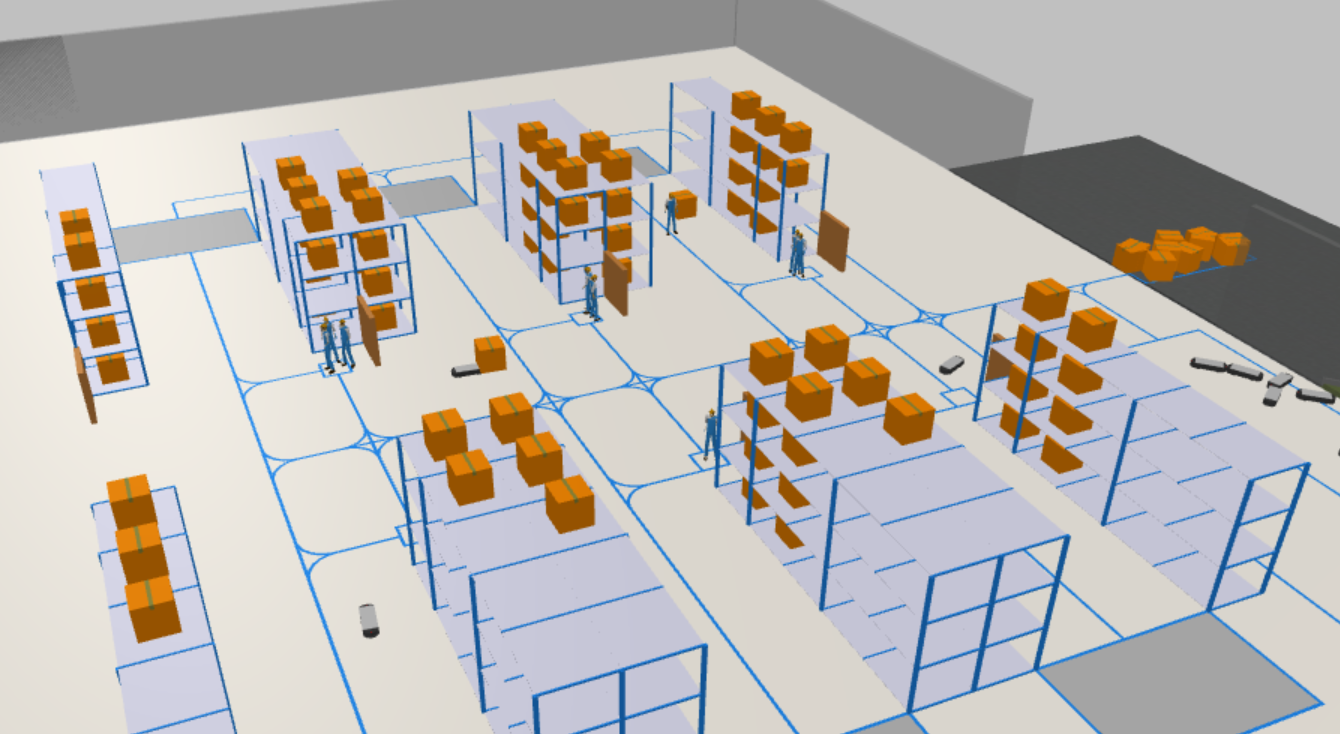}}
\caption{The simulation model of the use case.}
\label{fig_simulator}
\end{figure}

The DT simulation model of the use case is implemented using AnyLogic, as shown in Fig. \ref{fig_simulator}. In total 15 AMRs and 9 workers are involved. AMRs and workers both move 1 m/s at maximum. The safety compliance rule embedded in each AMR is: \textit{if there is any obstacle, AMR, or human worker within y meters, reduce the speed to 0.5 m/s}. Here \textit{y} is a parameter subject to change during runtime with DT simulation analysis.

The environment is designed to be dynamic, with orders arriving at varying rates throughout time.  Combined with the intrinsic complexity due to interaction and collaboration between workers and AMRs and the considered constraints for the given context (e.g., safety, spatial distance etc), the resulting productivity is significantly impacted. 

% How the ``dynamism'' of the environment is designed.
The experiment in this section considers a \textit{two-phase} scenario to illustrate such dynamism: in \textit{Phase 1}, the orders arrive every 50 seconds, which is considered a relatively slower rate for this case study; later in \textit{Phase 2}, the orders arrive at a much faster rate -- every 15 seconds. 
The two phases depict two typical scenarios.
In Phase 1, there is no need for the workers and AMRs to hurry to fulfil the order. However, in Phase 2, time is more stringent and more AMRs may be needed, which can lead to a more crowded warehouse and potentially unsafe conditions for human workers. This can result in a violation of safety compliance measures.  

\subsection{Phase Change}
\begin{figure}[!t]
\centerline{\includegraphics[width=\columnwidth]{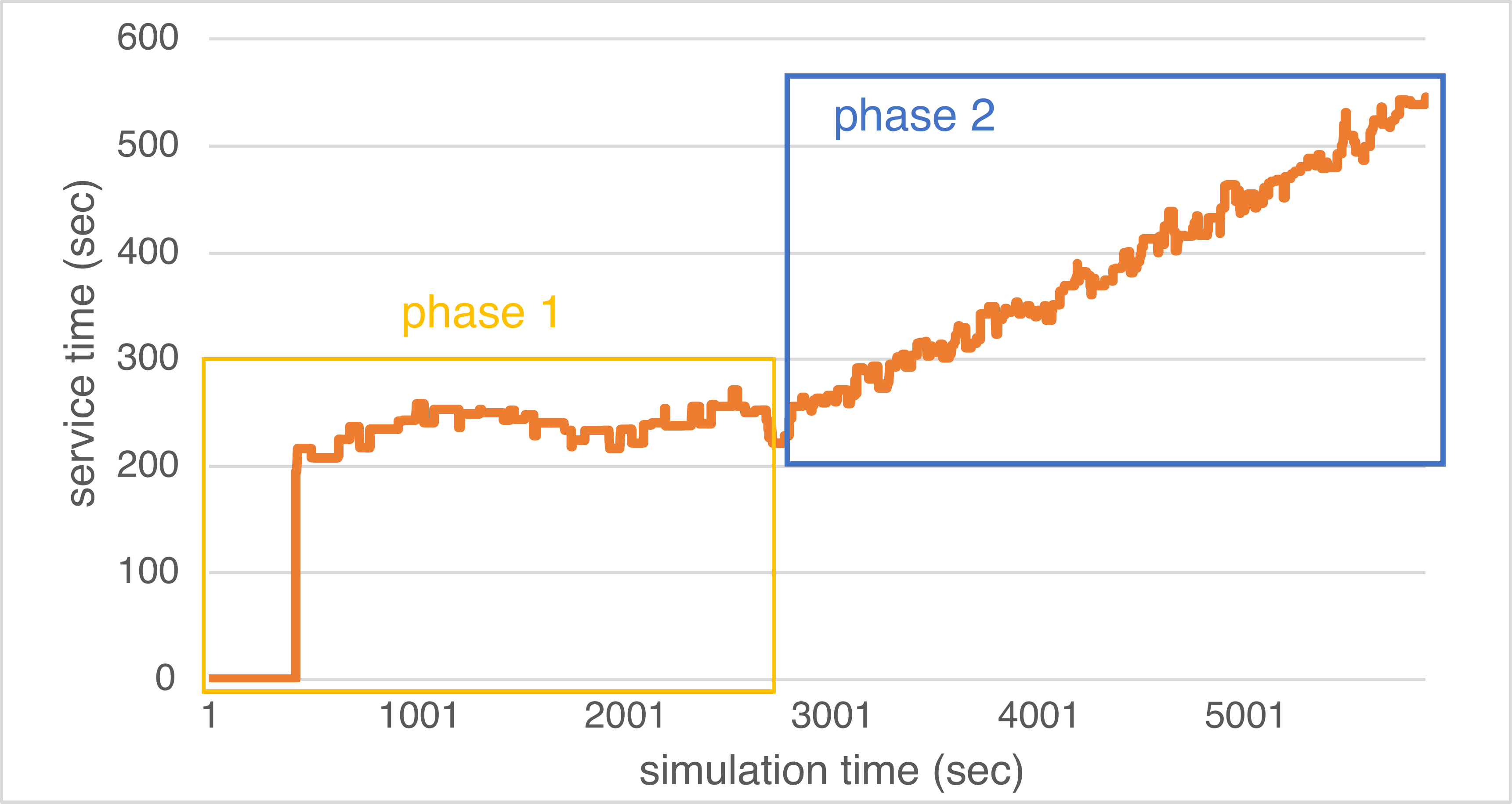}}
\caption{Productivity (service time) monitoring for the two-phase scenario with the rule parameter $y=5$. Phase 2 causes an increase in service time.}
\label{fig_monitoring}
\end{figure}

\begin{figure}[!t]
\centerline{\includegraphics[width=0.8\columnwidth]{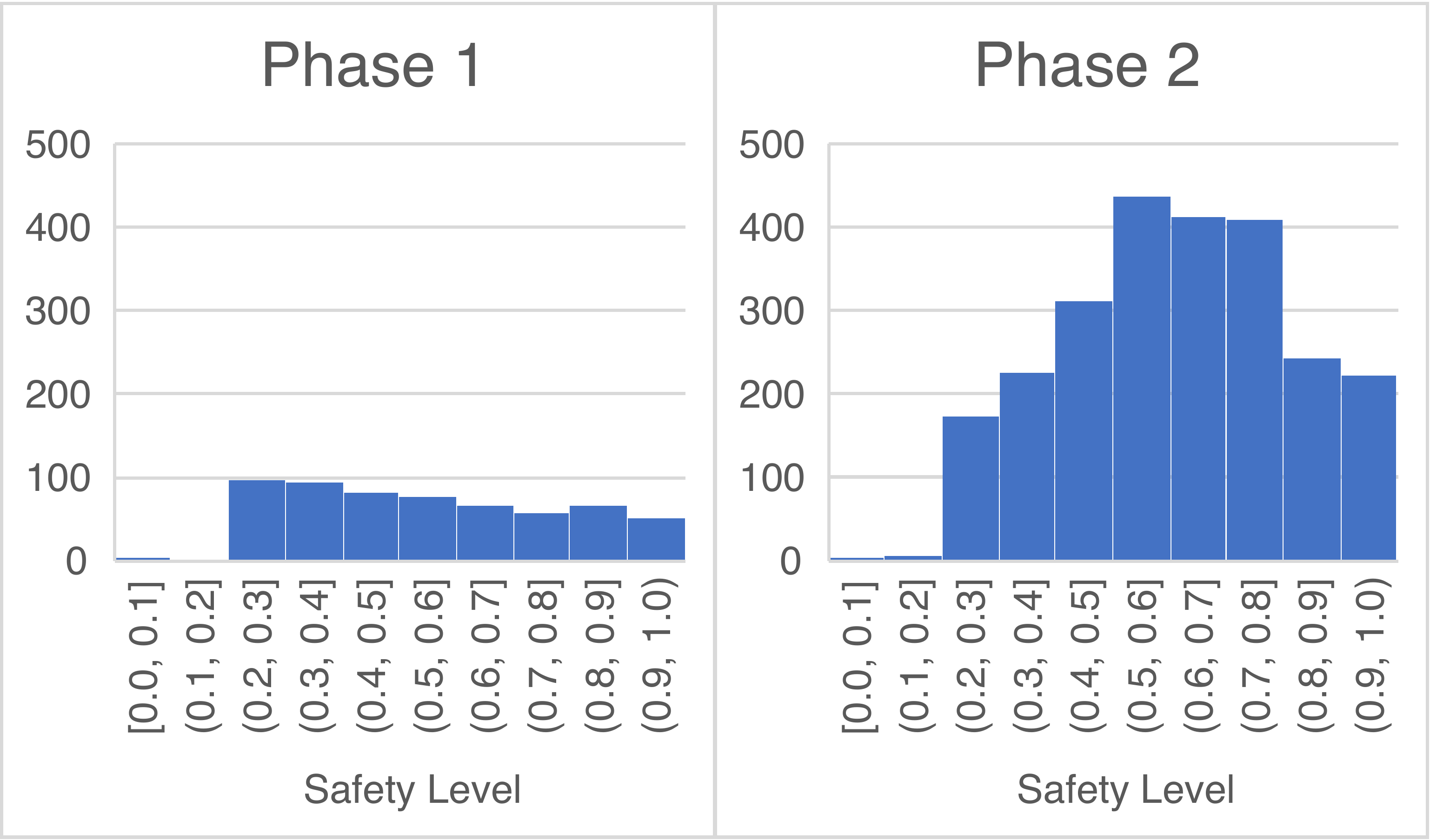}}
\caption{Histogram of the monitored overall safety level (measured by $\text{Safety}_{\text{min}}$ in equation (\ref{eqn: safety_min}) with threshold $d_{th}=4 \ \text{meters}$) through time in the two phases when the rule parameter $y = 5$., excluding the value of 1. The safety level in Phase 2 is more likely to be lower compared to that in Phase 1.}
\label{fig_histogram}
\end{figure}

% First, we use one simple case to show that:
% a rule acceptable in one phase may not be optimal in the later phase.
We first show that a rule leading to acceptable goal satisfaction in one phase may have poor performance in a later phase using a trivial parameter value.
The rule parameter \textit{y} of the AMRs is set to be 5 in both Phase 1 and Phase 2. The simulation result is shown in Fig. \ref{fig_monitoring} and Fig. \ref{fig_histogram}. 

During Phase 1, productivity remains at an acceptable level as the curve remains relatively flat, indicating that no tasks are accumulating. 
However, in Phase 2, following the same rule parameter results in a steady rise in service time, which indicates that the tasks cannot be completed promptly.
Furthermore, regarding the safety concerns depicted in Fig. \ref{fig_histogram}, it is evident that the safety level of workers in Phase 2 is lower than that of Phase 1. This is because workers tend to be closer to AMRs more frequently during Phase 2.
Therefore, according to the data, the rule must be adapted in Phase 2 to re-optimise the trade-off between productivity and safety. 

It can be argued that such knowledge can be obtained through offline testing during design time. However,
before deploying the system to production, it is challenging to predict the occurrence of a phase change and its characteristics. 
Offline analysis requires identifying all potential phases, and considering all possible world states when the phase change occurs. 
Due to the immense number of potential combinations of states and phases in such a multi-agent system, this exhaustive offline analysis is not feasible.

% It would be better if we could show how the performance will be when the same phase changes at different points in time (meaning: the same phase change, but at different states).

Nevertheless, with a DT,
compliance and safety engineers can use the DT to conduct \textit{runtime} what-if analysis for productivity vs. safety compliance, and to evaluate different alternative rule designs when a specific phase change is identified during runtime. 
The premise is that the exercise can help in better impromptu and ``directional'' planning and optimisation of the production environment specifically for the current state and phase, through adaption and/or tuning the parameters, first tested and evaluated in the simulation model during runtime, before it gets implemented into the physical system\footnote{Such analysis requires faster-than-real-time simulation for multi-agent systems, which can be achieved by various execution speedup approaches such as parallel and distributed simulation \cite{suryanarayanan_range_query_2010,logan_interest_management_2000}, but this is not the main focus of this paper.}.

\subsection{Runtime What-if Analysis}
\begin{figure}[!t]
\centerline{\includegraphics[width=0.9\columnwidth]{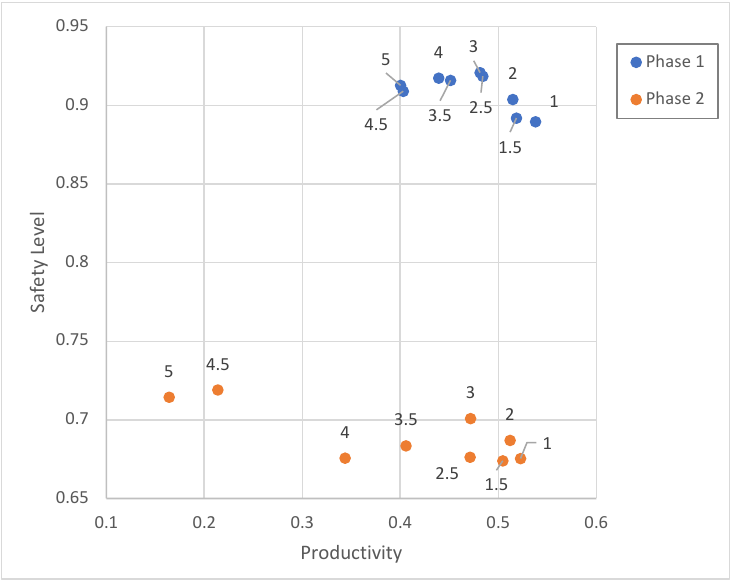}}
\caption{The safety and productivity metrics when different design alternatives (the choice of the rule's parameter value) are applied to Phases 1 and 2. Each data point represents the result of selecting a parameter value $y$. The data labels show the value of $y$ for each data point. The safety level is calculated according to equation (\ref{eqn: safety_min}) with $d_{th} = 4$ meters and the productivity is calculated with $T_{th} = 400$ seconds. }
\label{fig_pareto_result}
\end{figure}

We then evaluate the impact of runtime what-if analysis enabled by DT. We examine various design alternatives for the rule during Phases 1 and 2, and the simulation results are exhibited in Fig. \ref{fig_pareto_result}. The ideal solution should maximise both the safety level (safety compliance) and productivity.
% What we do:
% We explore many design alternatives and evaluate their outcome specific to each phase.
% 
% Why we do that:
% ???????
% 
% Observation: 
% - The optimal value (given a preference) is different in both phases
% - The sensitivity of the values is different in the two phases.
% - parameters that are optimal during certain phases or scenarios may not guarantee their performance later in the future. 
% 
% Indication: 
% - DT can efficiently obtain this knowledge compared to offline design. A more ``directed'' exploration compared to offline analysis.
% -- If we can run what-if analysis when we notice the phase changes, we can efficiently explore the design space that contributes specifically to that phase.
% -- Without DT monitoring, we must exhaustively explore all possible scenarios at design time, which is infeasible due to the large design space.
% - The DT can also support runtime changes in the preferences of the goals.

\begin{figure}[!t]
\centerline{\includegraphics[width=\columnwidth]{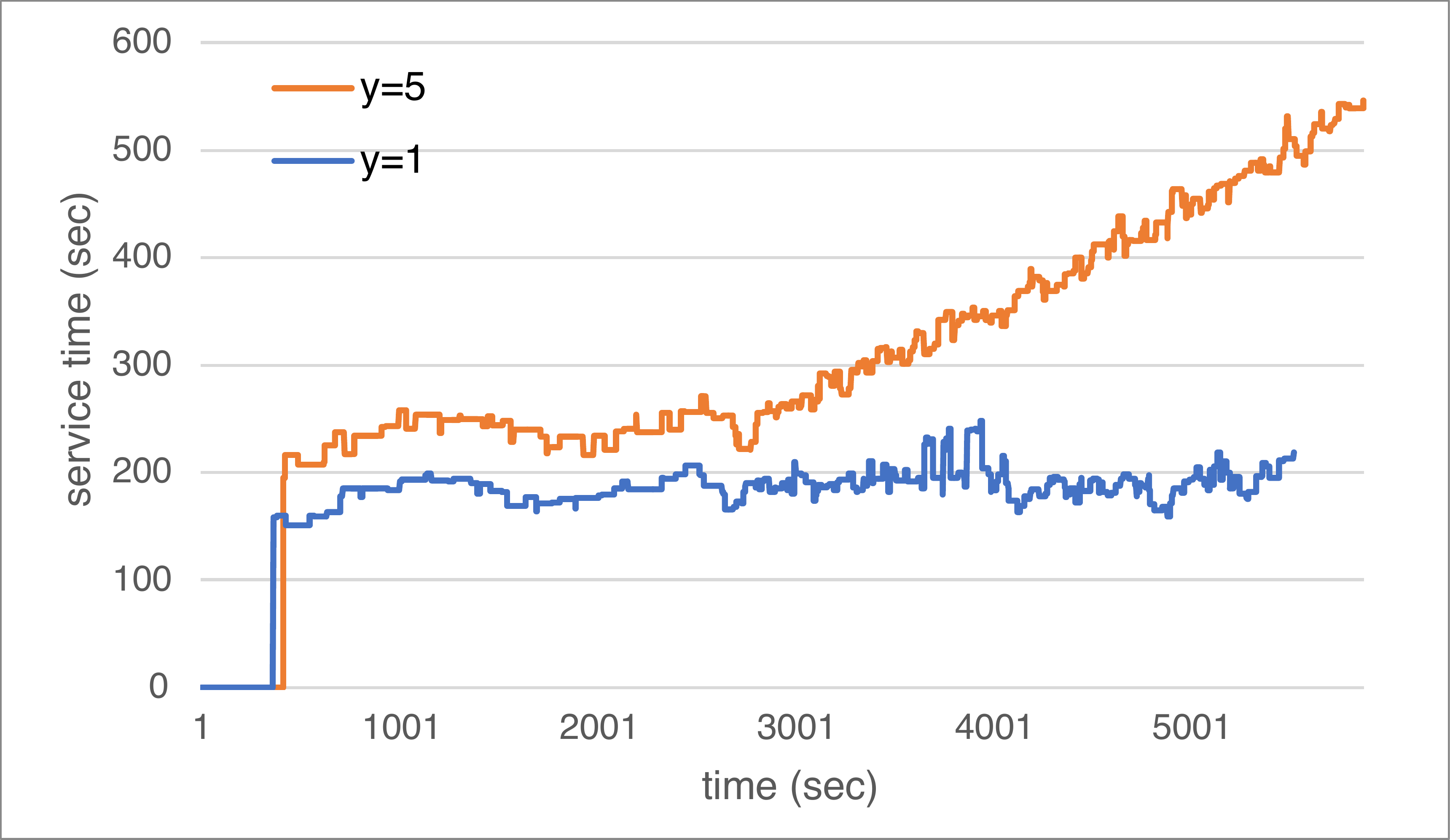}}
\caption{Comparison of two design alternatives in both Phase 1 and Phase 2.}
\label{fig_both}
\end{figure}

The optimal design alternative (parameter value of the rule) depends on the preference of the two objectives.
According to Fig. \ref{fig_pareto_result}, if maximising productivity is preferred, the optimal value is 1 for both phases. This is demonstrated in Fig. \ref{fig_both}, which displays lower service time. 
Similarly, if safety is preferred in both phases, then in Phase 1, the optimal value for $y$ is 3, but in Phase 2, $y$ should be changed to 4.5 to maximise the safety level.
In practice, preference is a human factor and thus is subject to change during the production process. Therefore, DT runtime analysis is highly required to incorporate variable preference into the assessment of design alternatives for the trade-off analysis.

Additionally, it can be observed that different phases have varying sensitivity to design alternatives.
Fig. \ref{fig_pareto_result} clearly indicates that Phase 2 is much more sensitive to parameter values than Phase 1.
Consider the comparison between values 3 and 4.5 as an example. During Phase 1, value 3 demonstrates greater Pareto efficiency due to its higher safety level and productivity. However, in Phase 2, both values are on the Pareto front, with value 3 exhibiting higher productivity and value 4.5 demonstrating a higher safety level. This serves as a reminder that parameters that prove optimal in certain phases or scenarios may not necessarily remain optimal in the future.
Therefore, it is imperative to understand that during Phase 2, proper adaptation of the rule parameter is crucial, as even the slightest differences in values can significantly impact productivity. Efficient acquisition of such knowledge is only viable with DT that conducts runtime analysis for the specific phase and state observed at runtime, as opposed to exhaustive offline testing of all possible phases and world states.

\subsection{Summary}

% What we can observe from the experiments
The above analysis shows that dynamic phase changes in the environment require re-evaluations of design alternatives to adapt for better trade-offs between compliance and quality goals.
However, such analysis is infeasible offline because of the unpredictable nature of the complex system and the resulting vast exploration space. 
It is difficult to predict when and what specific phase can occur under what world state; thus, offline analysis can only exhaustively explore all possibilities.

In contrast, 
DT can explore alternative designs more efficiently by obtaining knowledge specific to the phase change that is happening in production.
This results in a targeted exploration without simulating scenarios unrelated to the monitored reality.
DT also supports runtime \textit{dynamic} preference, an unpredictable human factor that is difficult to handle by offline analysis.
All the above capabilities are made possible through DT's monitoring ability and the multi-model approach with high-fidelity simulation models and goal models.

Compliance and safety engineers can use the approach in %either of the following modes: (i) assist in offline planning and then tune/update the physical system for what is optimal for either safety, productivity or both; or (ii) 
online adaption, where physical system and DT embark on frequent and continuous symbiotic updates, where the physical system is updated  with the optimal solution for a given context, after being tested in the DT.

\section{Conclusion} \label{sec:conclusion}
This paper proposes a novel Digital Twin (DT) approach for runtime compliance management, leveraging multi-agent simulation and goal modelling. 
The approach is demonstrated using a smart warehouse human-robot collaboration case with stringent safety compliance and productivity goals.  
Alternative rule designs of each agent are first elicited with GRL goal modelling for safety compliance and  productivity, along with other dependencies. Then these alternatives will be evaluated at runtime at the specific scenario encountered with the support of faster-than-real-time high-fidelity agent-based simulation. Finally, the rule leading to the ``best'' trade-off will be applied to the agent on the robot.
Through Pareto efficiency, the trade-off between compliance and productivity can be evaluated. 
Our DT-oriented architecture provides compliance engineers with forecasting capability for refining and switching between the different rules before enacting the real system at runtime. 
The experimental evaluation demonstrated the need for runtime adaptive compliance and the efficacy of using DT what-if simulation analysis for adaptive control.
It is shown that DT can more efficiently explore the design space and understand the rules in different contexts.
%It is shown that leveraging DT has the potential for continuous and live runtime adaptation of behavioural rules and parameters affecting the satisfaction of compliance. 
Our future work will examine how the vision can be extended to provide cognitive and proactive adaptive control considering various compliance requirements, context and time-varying trade-offs.

% ---------------
% Acknowledgement of financial support should be placed at the end of the first footnote.
% Ref1: https://www.ieeesmc.org/publications/transactions-on-smc-systems/information-for-authors/
% Ref2: https://www.ieeesmc.org/publications/transactions-on-cybernetics/information-for-authors-3/
% ---------------

%\section*{Acknowledgements}
% This research was supported by: Shenzhen Science and Technology Program,  China (No. GJHZ20210705141807022); SUSTech-University of Birmingham Collaborative PhD Programme; Guangdong Province Innovative and Entrepreneurial Team Programme, China (No. 2017ZT07X386); SUSTech Research Institute for Trustworthy Autonomous Systems, China; and EPSRC/EverythingConnected Network project on Novel Cognitive Digital Twins for Compliance, UK.

\bibliographystyle{IEEEtran}
\bibliography{references}

\end{document}